\documentclass[useAMS]{mn2e}

\usepackage{graphicx}
\usepackage{amssymb}

\title{On the Dispersion Measure of High-Redshift Synchrotron Sources}
\author[R.V.E. Lovelace \& D.W. Richards]{R.V.E. Lovelace$^{1}$
\& D.W. Richards$^{2}$\\ 
$^{1}$Department of Astronomy, Cornell University, Ithaca, NY
14853; email: lovelace@astro.cornell.edu\\
$^{2}$Department of Physics, University of South Florida, Tampa, FL 33612; email: drichard@health.usf.edu\\}

\begin{document}

\maketitle

\begin{abstract}

      A method is proposed for deriving 
the dispersion measure (related to the line-of-sight integrated free electron density) to high-redshift, powerful synchrotron radio sources by  analysis of the temporal/spectral properties of the recorded signals.      This would
allow direct measurements of the distributed density of ionized baryons
along the line-of-sights to the sources.

\end{abstract}

\begin{keywords}

cosmology: miscellaneous --- cosmology: observations
 
\end{keywords}

\section{Introduction}
    Highly dispersed
radio pulses of likely extragalactic origin 
have been detected at GHz frequencies  - one
with a dispersion measure  DM$=\int d\ell ~ n_e=375 $
pc(cm)$^{-3}$ (Lorimer et al. 2007) and
another with DM$=746$ pc(cm)$^{-3}$ (Keane et al. 2012).
  Here,  the
integral is from the earth to the source and $n_e$ 
is the free electron density.   
    Unfortunately the pulses
do not appear to repeat and the distances to their sources
are unknown.   Here, we propose that the dispersion
measures to high-redshift incoherent synchrotron radio
sources can be extracted by analysis of the amplitude
cross-correlation of the radio waves received at two nearby
frequencies.    This method utilizes the broadband nature
of  the synchrotron radiation.

   Section 2 discusses the theory first without intervening 
plasma and then including it.   Also, the influence of
interstellar/intergalactic scintillations are considered.
Section 3 gives our conclusions.

\section{Theory}

   The measured electric field on the ground
from a distant optically-thin incoherent synchotron source can be approximated as
\begin{equation}
E(t) = \sum_j {\cal E}_j \delta(t-t_j)~,
\end{equation}
where the different terms represent the contributions from the individual
electron or positron in the source, ${\cal E}_j$
is the complex field amplitude, and $\delta$ is the delta function.
   The synchrotron radiation from a highly relativistic electron with
Lorentz factor $\gamma \gg 1$  is directed in a narrow beam of
angular width $\sim 1/\gamma$ in the direction of the electron's instantaneous velocity (e.g., Pacholczyk 1970).  This narrow beam sweeps across the 
observer in a  time interval $dt \sim (\gamma^3 \omega_c)^{-1}$
which is much less than the gyration period of the electron,
$2\pi/\omega_c$.   Here, $\omega_c =eH/(m_e\gamma c)$ is
the relativistic cyclotron frequency with $e$ and $m_e$ the charge
and rest mass of the electron and $H$ the magnetic field strength in
Gauss.
     Thus the radiation at a point on the ground is in
narrow pulses each of width $\sim dt$ which is  the basis
for using delta functions in equation (1).
   Consequently the frequency spectrum of
the radiation is very broad with a maximum at $\omega_{\rm syn}
\sim (dt)^{-1} = \gamma^3 \omega_c$.    
    Because the synchrotron radiation is dominantly linearly
polarized,  we consider $E$ to be one of the linear polarization components.
 For the moment we neglect the influence of  intervening plasma.
       The representation of $E(t)$ by equation (1), where ${\cal E}_j$
has a broad frequency spectrum, is specific to  incoherent synchrotron radiation. 
    It does not apply for example to thermal black-body radiation or inverse Compton radiation.

        We consider that the signal (1) is first passed
through a bandpass filter which passes only 
frequencies in the interval $f\pm B/2$, where $B\ll f$.
The resulting signal is then  mixed
with a precise local oscillator  reference signal $\exp(2\pi {\rm i }f t)$ to give
\begin{equation}
\overline{E}(t) = \sum_j {\cal E}_j \Delta(t-t_j)~,
\end{equation}
where $\Delta(\tau) $ is the function which results from
passing the delta function through the bandwidth $B$.
For example, for a flat response within $B$,
$\Delta(\tau) =\sin (\pi B \tau)/(\pi \tau)$ which has a
temporal width $\sim  B^{-1}$.    
    The expansion (2) is similar that of Appendix
A of Cordes et al. (2004).

    The average power is proportional to an average
of $|\overline{E}(t)|^2$.     We can assume $\langle {\cal E}_j
{\cal E}_k^* \rangle = \langle|{\cal E}|^2\rangle
\delta_{jk}$ because the different electrons are 
uncorrelated.   The individual pulses at times $t_j$ 
are closely spaced with a number per unit time ${\cal N}$
so that we can integrate over $t_j$ to obtain
\begin{equation}
\langle|\overline{E}|^2\rangle = {\cal N}B \langle |{\cal E}_j|^2\rangle~.
\end{equation}
 We have used $\int_{-\infty}^{\infty} dt [\Delta(t)]^2 =B$
 which holds for  $\Delta(t) =\sin (\pi B t)/(\pi t)$.
 
 Similarly,  the autocorrelation functions is found to be
\begin{equation}
{\langle \overline{E}(t)\overline{E}^*(t+\delta t)\rangle 
\over \langle |\overline{E}|^2\rangle} =
{1\over B}\int_{-\infty}^\infty dt \Delta(t)\Delta(t+\delta t)\equiv \rho_E(\delta t)~,
\end{equation} 
where the asterisk indicates the complex conjugate.
Note that $\rho_E(0) =1$ and that its width or correlation time is
$\sim B^{-1}$.   For the case $\Delta(t) =\sin (\pi B t)/(\pi t)$ one finds that $\rho_E( t) =\Delta( t)/B$.

    Consider the cross-correlation between the fields $\overline{E}_1$
and $\overline{E}_2$ at two separated frequencies $f_1$ and $f_2$.  
    The signal $\overline{E}_2(t)$ is obtained in a way similar to
that for $\overline{E}_1(t)$, but it is mixed with a second local
oscillator  signal
$\exp(2\pi {\rm  i }f_2 t)$, where the frequency $f_2$ is accurately locked
to $f_1$.
For simplicity we
assume both bandwidths  are equal to $B$.
    If the two frequencies are close together, $(f_2-f_1)^2 \ll f_1f_2$,
the amplitudes ${\cal E}_j$ do not change
significantly between the two frequencies because of
the broadband nature of the synchrotron radiation.
    Thus
\begin{equation}
{\langle \overline{E}_1(t)\overline{E}_2^*(t+\delta t)\rangle 
\over\left( \langle |\overline{E}_1|^2\rangle
\langle |\overline{E}_2|^2\rangle\right)^{1/2}} \approx
\rho_E(\delta t)~,
\end{equation} 
with $\rho_E$ given in equation (4).

   For the  cross-correlation obtained by time-averaging over 
time intervals $T$, the variation in the difference
between the two local oscillator frequencies, $d(f_2-f_1)$,
must satisfy
\begin{equation}
d(f_2-f_1) < (2\pi T)^{-1}~,
\end{equation}
 in order for the
correlation not to be smeared out.  
 Thus the local oscillators will need to
have very high stability possible with hydrogen
maser sources.

\subsection{Intervening Plasma: Dispersion}

     Propagation of the radio waves from a distant
high redshift source such as a quasar will modify
the cross-correlation because of the frequency
dependent delay time delay through the ionized
intergalactic plasma.  Ioka (2003)  and Inoue (2004)
discuss this intergalactic delay taking into account the redshifiting of 
the frequency and the time dilation during propagation
through a homogeneously  distributed fully ionized
plasma  assuming
a flat cosmology 
and express the delay as
\begin{equation}
\tau(f) ={e^2 \over 2\pi m_e c^2 f^2} {\rm DM}~,
\end{equation}
where DM is the dispersion measure,
\begin{equation}
{\rm DM} =\int_0^z dz^\prime
c\left |{dt \over dz^\prime}\right| {n_e(z^\prime) \over 1+z^\prime}~,
\end{equation}
where $|dt/dz| =[(1+z)H(z)|^{-1}$ and 
$H(z)=[\Omega_m(1+z)^3 +\Omega_\Lambda]^{1/2}$.
Assuming the plasma is without loss or gain
over time, the electron density varies as
 $n_e(z)=n_{e0}(1+z)^3$ with $n_{e0}=$ const.
Numerically,
\begin{equation}
\tau(f)  \approx 4.14\times 10^{-5} {DM \over f_{10}^2}~{\rm s}~,
\end{equation}
with DM in units of pc(cm)$^{-3}$ and
$f_{10}$ is the frequency in  units of $10^{10}$ Hz.
For redshifts $z$ small compared with unity
   ${\rm DM} = \int d\ell n_e$ where $d\ell$ is the path
element to the source.

   Ioka (2003) and Inoue (2004) have evaluated the
redshift integral (7) for a homogenous intergalactic medium 
and estimate that
${\rm DM} \approx1200 z $ pc(cm)$^{-3}$ 
for $z\lesssim 6$. 
    Additionally, there may  contributions to DM
due to propagation through the
free electrons of our Galaxy and/or the source galaxy.
For our galaxy estimates of the DM in directions perpendicular/parallel
to the disc plane range from $30$ to $1000$pc(cm)$^{-3}$
(Taylor et al. 1993).

        In order for the pulse-like nature of the synchrotron
radiation not be smeared out by the variation of $\tau(f)$ across
the  receiver bandwidth $B$, we must have  $B|d\tau/df| <
B^{-1}$.  This is the same as the requirement
\begin{equation}
{B } < 1.1\times 10^{7} {f_{10}^{3/2} \over {\rm DM}^{1/2}}{\rm Hz} 
\approx 3.16\times10^{5}{f_{10}^{3/2}\over z^{1/2}}~{\rm Hz}~,
\end{equation}
where the last expression uses DM $\approx 1200z$.
   We assume that $B$ satisfies this condition.

   Consider again the cross-correlation
of the fields at two nearby frequencies, $f_1$ and $f_2=f_1+\delta f$,
again with equal bandwidths $B$.
    We have in place of equation (5), 
\begin{equation}
{\langle \overline{E}_f(t-\tau_f)\overline{E}_{f+\delta f}^*(t+\delta t-\tau_{f+\delta f})\rangle 
\over \langle |\overline{E}_f|^2\rangle} =
\rho_E(\delta t-\tau_f^\prime \delta f)~.
\end{equation}
Here, $\tau_{f,f+\delta f}=\tau(f,f+\delta f)$ and $\tau_f^\prime =d\tau(f)/df$. 
    The maximum of $\rho_E$ occurs at $\delta t = \tau_f^\prime \delta f$.   
Thus measurements  of the values $(\delta t_m,~\delta f_m)$ of the maximum
allows  one to derive $\tau_f^\prime =\delta t_m/\delta f_m$ which
is directly related to the dispersion measure.

    From equation (9) we have
\begin{equation}
\tau_f^\prime =-8.28\times 10^{-15} {{\rm DM}\over f_{10}^3}\approx-10^{-11}{z\over f_{10}^3}~{\rm s}^2~.
\end{equation}
         In order for the shift of the cross-correlation
to be detectable with dispersion included we need $|\tau_f^\prime|\delta f > B^{-1}$.
   In view of inequality (10) this implies that we need
\begin{equation}
\delta f > 1.1\times 10^7{f_{10}^{3/2} \over {\rm {\rm DM}}^{1/2}}
\approx 3.16\times 10^5 {f_{10}^{3/2} \over z^{1/2}}~{\rm Hz}~,
\end{equation}
which is the reverse of inequality (10).

      The small bandwidths indicated by inequality (10)
may lead to poor signal to noise ratios in
measurement of the cross-correlation functions.
This may be offset by use of a filter bank with
a large number $N\gg 1$ of uniformly spaced channels
each  ($j=1..N$) of bandwidth $B$.   
   The different channels can be averaged to give the cross-correlation as
as $\sum_j \langle\overline{E}_j(t-\tau_f) \overline{E}_{j+k}^*(t+\delta t-\tau_{f+\delta f})\rangle$ where $j$ corresponds to $f$ and $k$ to $\delta f=$const.
   Techniques and software for extracting weak dispersed pulses from
radio measurements have been highly developed for applications to
pulsar searches (Ransom, Eikenberry, \& Middleditch 2002; Ransom 2012).

    A test and calibration of the proposed method can be
obtained by applying it to the strong synchrotron emission of the Crab
Nebula because in this case the dispersion measure is
known from measurements of the Crab pulsar.  It is DM $\approx 56.77$
pc(cm)$^{-3}$ (Lundgren et al. 1995).   Inequality (7) then gives
$B<1.46f_{10}^{3/2}$MHz.  The time delay in the cross-correlation
between two frequencies separated by say $\delta f =5B$ is 
$\delta t = -3.44\mu$s.

\subsection{Interstellar/Intergalactic Scintillations}

    The interstellar scintillations (ISS) of small
angular diameter extragalactic sources has been
discussed by Lovelace \& Backer (1972) and Goodman (1997).
    Lovelace and Backer  considered the random modulation in the frequency domain of a wave from a broadband point source caused by the
 propagation through electron density irregularities in the interstellar medium (ISM).   
      There is in general a ``frequency
correlation scale,'' $\Delta f$, determined by the properties
of the ISM.    
     Two frequencies components of a wave from a point source
separated by more than $\Delta f$ are  uncorrelated due to
the random multi-path propagation.   That is, the phase shift
due to the extra propagation path length $\pi f\Theta_s^2 Z/c$
differs by more than $2\pi$ for frequencies separated by
more than $\Delta f$, where $\Theta_s$ is the scattering angle
(proportional to $f^{-2}$)
and $Z$ is the effective distance to the scattering screen.
    The value of $\Delta f$ depends
on the radio frequency and the propagation distance through
the ISM.   
     At frequencies $f\lesssim f_*\approx 2$ GHz, 
the scintillations are strong and $\Delta f/f \sim
(f/f_*)^3$, whereas at higher frequencies the
scintillations are weak and $\Delta f/f \sim 1$
(Lovelace \& Backer 1972).    Lovelace and
Backer  also estimated the   critical angular size $\psi$ of extragalactic
radio sources to be of the order of $25\times 10^{-6}$ arc sec. 
Sources of smaller size may exhibit random intensity variations whereas
larger sources are steady.   
   The calculated shift in the  cross correlation function  arises from the contribution of the individual radiating particles
in the source  and is not affected by the source size being larger or smaller than $\psi$.

      Scintillations due to propagation through the inhomogeneous
intergalactic medium are considered by Goodman (1997) and
estimated to be small compared with the ISS.

    In order for the shift in the cross-correlation function (equation 8)
to be measurable we must have $ \delta f < \Delta f$.   This can
be easily satisfied at GHz frequencies.

\section{Conclusions}   

   The detection of highly dispersed radio pulses
likely of extragalactic origin
by Lorimer et al. (2007) and Keane et al. (2012) 
opens the intriguing possibility that the sources
of such pulses can be identified and thus their
distances determined.   
    This would allow direct measurements
of the distributed baryon density of the intergalactic medium along
lines-of-sight to possibly high redshift sources.
This is of evident cosmological interest.    
   Unfortunately,
the narrow, highly dispersed radio pulses appear to be rare
and non-repeating.   Hence  the distances
to the sources are remain unknown.

    More recently, a search by Bannister et al. (2012) for prompt radio pulses at $1.4$GHz from gamma ray burst sources - some at cosmological
distances -  has revealed highly dispersed pulses.  However,
the association of the pulses with the gamma ray burst
sources remains uncertain.

   This work proposes a method to measure the dispersion
measures of high-redshift incoherent synchrotron radio
sources associated with quasars  by analysis of the amplitude
cross-correlation of the radio waves received at two nearby
frequencies.        
    This method would  allow direct measurements
of the distributed baryon density of the intergalactic medium along
lines-of-sight to sources of known high redshifts.
      This method utilizes the broadband nature
of  the synchrotron radiation.   The bandwidths needed
by this method are much narrower  
 than those used to detect the mentioned dispersed radio pulses.
    Two key requirements on this method are identified: 
 one is that the local
oscillators at the two frequencies  have very high stability
(inequality 6) and the second is that the bandwidths be sufficiently narrow
(inequality 10).

\section*{Acknowledgments}

    We thank  Shami Chaterjee, Jim Cordes,
Don Campbell, Phil Kronberg, and Alex Pidwerbetsky for valuable comments.
We thank two referees, one Jean-Pierre Macquart and the
other anonymous, for valuable criticism of an earlier
version of this work.

\end{document}